\documentstyle{article}

\begin{document}
\baselineskip=18pt
\title{A $\hbar$-deformation of the $W_{N}$ algebra and its vertex operators }

\author{
 Bo-yu Hou$^{a}$ 
\thanks{e-mail address:byhou@nwu.edu.cn}
\hskip 0.5truecm Wen-li Yang$^{a,b}$ \\
\bigskip\\
$^{b}$ CCAST (World Laboratory), P.O.Box 8730, Beijing 100080, China\\
$^{a}$ Institute of Modern Physics, Northwest 
University, Xian 710069, China
\thanks{Mailing address
}}
\maketitle

\begin{abstract}
In this paper,we derive a $\hbar$-deformation of the $W_{N}$ algebra and its 
quantum Miura tranformation.The vertex operators for this $\hbar$-deformed  
$W_{N}$ algebra and its commutation relations are also obtained.
\end{abstract}

\section{Introduction}
Recently,the studies of q-deformation of some 
infinite dimensional algebra---q-deformed affine
algebra[4,6,11] ,q-deformed virasoro[2,21] algebra and $W_{N}$-algebra
[2,3,9,10],have attracted  much attention of physicist and mathematician.
The q-deformed affine algebra 
and its vertex operators provide a powerful method to study the state space 
and the correlation function of solvable lattice model both in the bulk [16] 
case and the boundary case[14].However,the symmetry of q-deformed affine 
algebra only correspond to the current algebra ( affine Lie algebra) 
symmetry in the conformal field theory (CFT),not to the Virasoro 
and W algebra 
type symmetry.The q-deformation of Virasoro and W algebra ,which would play 
the role of symmetry algebra for the solvable lattice model,has been 
expected for a long time. H.Awata et al also constructed the q-deformed 
$W_{N}$ algebra (including Virasoro algebra) and associated Miura 
transformation from studying the Macdonald symmetrical functions[2].
On the other hand,E.Frenkel and N.Reshetikin succeeded in 
constructing the q-deformed classical $W_{N}$ algebra and corresponding Miura 
transformation in analysis of the $U_{q}(\hat{sl}_{N})$ algebra at the 
critical level[10].Then,B.Feigin and E.Frenkel obtained the quantum 
version of  
this q-deformed classical $W_{N}$ algebra,i.e the q-deformed $W_{N}$ 
algebra[9].The q-deformed Virasoro algebra was also given by 
S.Lukyanov et al 
in studying the Bosonization for ABF model[21].The bosonization for vertex 
operators of q-deformed Virasoro[17,21] and $W_{N}$ algebra[2,1] have been 
constructed.  

However,there exists another important deformation of infinite dimensional 
algebra,which plays an important role in the completely integrable field 
theories(In order to comparation with q-deformation, we call it as 
$\hbar$-deformation).This deformation for affine algebra was originated by  
Drinfeld in studies of Yangian[7].It has been shown that the Yangian 
($DY(\hat{sl}_{2})$) is the dynamical non-abelian symmetry algebra for 
SU(2)-invariant Thirring model[13,18,19,23].Naively,
the $\hbar$-deformed affine algebra  
(or Yangian) would play the same role in the integrable field theories as the 
q-deformed affine algebra in solvable lattice model.Naturally,the 
$\hbar$-deformed Virasoro and W algebra,which would play the role of symmetry  
algebra of some integrable field model,are expected.We have succeeded in 
constructed $\hbar$-deformed Virasoro algebra in Ref.[12] and shown that  
this $\hbar$-deformed Virasoro algebra is the dynamical symmetry algebra of 
the Restriected sine-Gordon model.In this paper,we construct the 
$\hbar$-deformed $W_{N}$ alegbra (including the $\hbar$-deformed Virasoro 
algebra as a special case of N=2) , corresponding quantum Miura 
transformation and its vertex operators . The $\hbar$-deformed $W_{N}$ 
algebra will become the usual no-deformed $W_{N}$ algebra[8] with some  
center charge which is related to parameter $\xi$ ,when 
$\hbar \longrightarrow 0$ and $\xi$ and $\beta$ are fixed. 

This paper is arranged as follows: In section 2,we define the 
$\hbar$-deformed $W_{N}$ algebra and its Miura transformation.The screening  
currents and vertex operators are derived in section 3 and 4.

\section{$\hbar$-deformation of $W_{N}$ algebra}
In this section,we start with defining a $\hbar$-deformed $W_{N}$ algebra 
by quantum Miura transformation.

\subsection{$A_{N-1}^{(1)}$ type weight}

In this subsection, we shall give some notation about $A_{N-1}^{(1)}$ 
type weight which will be used in the following parts of this paper.
Let $\epsilon_{\mu}(1\leq \mu \leq N )$ be the orthonormal basis in 
${\bf R}^{N}$,which is supplied with the inner product 
 $<\epsilon_{\mu},\epsilon_{\nu}>=\delta_{\mu \nu}$.
Set 
\begin{eqnarray}
& &\overline{\epsilon}_{\mu}=\epsilon_{\mu}-\epsilon\ \ \ \,\ \ \ 
\epsilon=\frac{1}{N}\sum_{\mu=1}^{N}\epsilon_{\mu}
\end{eqnarray}
The $A^{(1)}_{N-1}$ type weight lattice is the linear space of the 
\begin{eqnarray*}
P=\sum_{\mu=1}^{N}Z\overline{\epsilon}_{\mu}
\end{eqnarray*}
Note that $\sum_{\mu=1}^{N}\overline{\epsilon}_{\mu}=0$.Let $\omega_{\mu}
(1\leq\mu\leq N-1 )$ be the fundamental weights:
\begin{eqnarray*}
\omega_{\mu}=\sum^{\mu}_{\nu=1}\overline{\epsilon}_{\nu}
\end{eqnarray*}
\noindent and the simple roots $\alpha_{\mu}(1\leq\mu\leq N-1)$:
\begin{eqnarray}
\alpha_{\mu}=\overline{\epsilon}_{\mu}-\overline{\epsilon}_{\mu +1}
=\epsilon_{\mu}-\epsilon_{\mu +1}
\end{eqnarray}
An ordered pair $(b,a)\in {\bf P}^{2}$ is called admissible if only if 
there exists $\mu \in (1\leq\mu\leq N-1)$ such that 
\begin{eqnarray*}
b-a=\overline{\epsilon}_{\mu}
\end{eqnarray*}
An ordered set of four weight $\left(\begin{array}{ll}c&d\\b&a
\end{array}\right)\in {\bf P}^{4}$ is called an admissible configuration 
around a face if and only if the pairs (b,a),(c,b),(d,a) and (c,d) are all  
admissible pairs.To each admissible configuration around a face we shall  
associate a Boltzmann weight in section 4.

\subsection{Quantum Miura transformation}

Let us consider free bosons $\lambda_{i}(t)\ \ (i=1,....,N)$ with continuous 
parameter $t\in \{ {\bf R}-0)\}$ which satisfy 
\begin{eqnarray}
& &[\lambda_{i}(t),\lambda_{i}(t')]=\frac{4sh\frac{(N-1)\hbar t}{2}
sh\frac{\hbar\xi t}{2}sh\frac{\hbar(\xi +1)t}{2}}{tsh\frac{N\hbar t}{2}}
\delta(t+t')\\
& &[\lambda_{i}(t),\lambda_{j}(t')]=\frac{4sh\frac{\hbar t}{2}
sh\frac{\hbar\xi t}{2}sh\frac{\hbar(\xi +1)t}{2}
e^{{\rm sign}(j-i)\frac{N\hbar t}{2}}}{tsh\frac{N\hbar t}{2}}
\delta(t+t')\ \ \ i\neq j
\end{eqnarray}
\noindent with the deformed parameter $\hbar$ and a generic parameter $\xi$ 
and $\lambda_{i}(t) $ should be subject to the following condition  
\begin{eqnarray}
\sum_{l=1}^{N}\lambda_{l}(t)e^{l\hbar t}=0
\end{eqnarray}
\noindent One can check that the restricted condition would be compatable 
with Eq.(3) and Eq.(4) .

\noindent {\bf Remark}: The free bosons with continuous parameter in the case 
of N=2 ,was first 
introduced by M.Jimbo et al  in studying Massless XXZ mode[15].This kind 
bosons could be used to construct the bosonization of Yangian double with 
center $DY(\hat{sl}_{N})$.

Let us define the fundamental operators $\Lambda_{i}(\beta)$ and the 
$\hbar$-deformed $W_{N}$ algebra generators $T_{i}(\beta)$ for i=1,...,N  
as follows
\begin{eqnarray}
& &
\Lambda_{i}(\beta)=:exp\{-\int^{\infty}_{-\infty}\lambda_{i}(t)
e^{i\beta t}dt\}:\\
& &T_{l}(\beta)=\sum_{1\leq j_{1}<j_{2}...<j_{l}\leq N}:
\Lambda_{j_{1}}(\beta+i\frac{l-1}{2}\hbar)
\Lambda_{j_{2}}(\beta+i\frac{l-3}{2}\hbar)....
\Lambda_{j_{l}}(\beta-i\frac{l-1}{2}\hbar):
\end{eqnarray}
\noindent and $T_{0}(\beta)=1$. Here $:O:$ stands for the usual bosonic 
normal ordering of some operator $O$ such that the bosons $\lambda_{i}(t)$ 
with non-negative mode $t>0$ are in the right. The restricted condition for 
bosons $\lambda_{i}(t)$ in Eq.(5) results in $T_{N}(\beta)=1$.Actually,the 
generators $T_{i}(\beta)$ are obtained by the following quantum Miura 
transformation 
\begin{eqnarray}
& &:(e^{i\hbar\partial_{\beta}}-\Lambda_{1}(\beta))
(e^{i\hbar\partial_{\beta}}-\Lambda_{2}(\beta-i\hbar))....
(e^{i\hbar\partial_{\beta}}-\Lambda_{N}(\beta-i(N-1)\hbar)):\nonumber\\
& &\ \ \ \ =\sum^{N}_{l=0}(-1)^{l}T_{l}(\beta-i\frac{l-1}{2}\hbar)
e^{i(N-l)\hbar\partial_{\beta}}
\end{eqnarray}
\noindent {\bf Remark}: $e^{i\hbar\partial_{\beta}}$ is the 
$\hbar$-shift operator such  that 
\begin{eqnarray*}
e^{i\hbar\partial_{\beta}}f(\beta)=f(\beta+i\hbar)
\end{eqnarray*}

If we take the limit of $\xi\longrightarrow -1$, the above generators 
$T_{l}(\beta)$ reduce to the classical version of $\hbar$-deformed 
$W_{N}$ algebra,whcih can be obtained by studying the Yangian Double 
with center $DY(\hat{sl}_{N})$ at the critical level (i.e l=-2).For 
the case of N=2, the corresponding classical $\hbar$-deformed $W_{2}$ 
(Virasoro) algebra has been given by Hou et al [5].Thus,we call the limit 
$(\xi\longrightarrow -1 $ with $\hbar$,$\beta$ fixed) 
as the classical limit.

Let us consider another limit:$\hbar\longrightarrow 0$ with fixed $\xi$ .
Then we have :$\Lambda_{i}(\beta)=1+i\hbar\chi_{i}(\beta)+o(\hbar)$ and 
$e^{i\hbar\partial_{\beta}}=1+i\hbar\partial_{\beta}+o(\hbar)$. Hence the 
right hand side of (8) in this limit becomes 
\begin{eqnarray}
:(i\hbar)^{N}(\partial_{\beta}-\chi_{1}(\beta))
(\partial_{\beta}-\chi_{2}(\beta)).....(\partial_{\beta}-\chi_{N}(\beta)):
+o(\hbar^{N})
\end{eqnarray}
\noindent and we obtain the normally ordered Miura transformation 
corresponding to the non-deformed $W_{N}$ algebra introduced by  
V.Fateev and S.Lukyanov[8].Therefore,the no-deformed $W_{N}$ algebra 
(the ordinary one ) with the center charge $(N-1)-\frac{N(N+1)}{\xi(1+\xi)}$ 
can be obtained by taking this kind limit.In this sense, we call this limit 
($\hbar\longrightarrow 0$ with fixed $\xi$ and $\beta$) as the 
conformal limit.
\subsection{Relations of $\hbar$-deformed $W_{N}$ algebra}
In order to get the commuting relations for bosonic operators, we should 
give a comment about regularization :When one compute the exchange relation 
of bosonic operators ,one often encounter an integral as follows 
\begin{eqnarray*}
\int^{\infty}_{0}F(t)dt
\end{eqnarray*}
\noindent which is divergent at $t=0$ .Hence we adopt the regularization 
given by Jimbo et al[15]. Namely, the above integral should be understood as   
the contour integral
\begin{eqnarray}
\int_{C}F(t)\frac{log(-t)}{2i\pi}dt
\end{eqnarray}
\noindent where the contour $C$ is chosen as the same as in the Ref.[15].
From the definition of fundamental operators $\Lambda_{i}(\beta)$ and the 
the commutation relations of bosons $\lambda_{i}(t)$ ,we can derive the 
following OPEs 
\begin{eqnarray}
& &\Lambda_{i}(\beta_{1})\Lambda_{i}(\beta_{2})= \phi_{i=i}
(\beta_{2}-\beta_{1}):\Lambda_{i}(\beta_{1})\Lambda_{i}(\beta_{2}):\\
& &\Lambda_{i}(\beta_{1})\Lambda_{j}(\beta_{2})= \phi_{i<j}
(\beta_{2}-\beta_{1}):\Lambda_{i}(\beta_{1})
\Lambda_{i}(\beta_{2}):\ \ \ i<j \\
& &\Lambda_{i}(\beta_{1})\Lambda_{j}(\beta_{2})= \phi_{i>j}
(\beta_{2}-\beta_{1}):\Lambda_{i}(\beta_{1})
\Lambda_{j}(\beta_{2}):\ \ \  i>j\\
& &\phi_{i=i}(\beta)=\frac{
\Gamma(\frac{i\beta}{N\hbar}-\frac{\xi}{N})
\Gamma(\frac{i\beta}{N\hbar}+1-\frac{1}{N})
\Gamma(\frac{i\beta}{N\hbar}+\frac{1+\xi}{N})
\Gamma(\frac{i\beta}{N\hbar}+1)}
{\Gamma(\frac{i\beta}{N\hbar})
\Gamma(\frac{i\beta}{N\hbar}-\frac{1+\xi}{N}+1)
\Gamma(\frac{i\beta}{N\hbar}+\frac{1}{N})
\Gamma(\frac{i\beta}{N\hbar}+1+\frac{\xi}{N})}\nonumber\\
& &\phi_{i<j}(\beta)=\frac{
\Gamma(\frac{i\beta}{N\hbar}-\frac{1}{N})
\Gamma(\frac{i\beta}{N\hbar}-\frac{\xi}{N})
\Gamma(\frac{i\beta}{N\hbar}+\frac{1+\xi}{N})}
{\Gamma(\frac{i\beta}{N\hbar}-\frac{1+\xi}{N})
\Gamma(\frac{i\beta}{N\hbar}+\frac{\xi}{N})
\Gamma(\frac{i\beta}{N\hbar}+\frac{1}{N})}\nonumber\\
& &\phi_{i>j}(\beta)=\frac{
\Gamma(\frac{i\beta}{N\hbar}+1-\frac{1}{N})
\Gamma(\frac{i\beta}{N\hbar}+1-\frac{\xi}{N})
\Gamma(\frac{i\beta}{N\hbar}+1+\frac{1+\xi}{N})}
{\Gamma(\frac{i\beta}{N\hbar}+1-\frac{1+\xi}{N})
\Gamma(\frac{i\beta}{N\hbar}+1+\frac{\xi}{N})
\Gamma(\frac{i\beta}{N\hbar}+1+\frac{1}{N})}
\end{eqnarray}
\noindent To caculate the general OPEs,the integral representation for 
$\Gamma$-function is very useful
\begin{eqnarray}
\Gamma(z)=exp\{\int^{\infty}_{0}(\frac{e^{-zt}-e^{-t}}{1-e^{-t}}+
(z-1)e^{-t})
\frac{dt}{t}\}\ \ \ \ ,\ \ \ Re(z)>0
\end{eqnarray}

\noindent {\bf Remark}: The above OPEs can be considered as the operator 
scaling 
limit of q-deformed $\Lambda_{i}(z)$ given by B.Feigin and E.Frenkel in 
studying the q-deformed $W_{N}$ algebra [9](the very similar 
$\Lambda_{i}(z)$ was also constructed by H.Awata et al [2]).The scaling 
limit is taken as following way
\begin{eqnarray}
z=p^{\frac{-i\beta}{\hbar}}\ \ \ ,\ \ q=p^{-\xi}\ \ ,\ \ 
\Lambda_{i}(\beta)=\lim_{p\longrightarrow 1}\Lambda_{i}(z)\equiv
\lim_{p\longrightarrow 1}\Lambda_{i}(p^{\frac{-i\beta}{\hbar}})
\end{eqnarray}

{\bf Theorem} 1 The generators $T_{1}(\beta)$ and $T_{m}(\beta)$ of 
$\hbar$-deformed $W_{N}$ algebra satisfy the following relations
\begin{eqnarray}
& &f_{1m}^{-1}(\beta_{2}-\beta_{1})T_{1}(\beta_{1})T_{m}(\beta_{2})
-f_{1m}^{-1}(\beta_{1}-\beta_{2})T_{m}(\beta_{2})T_{1}(\beta_{1})
=-2i\pi \{i\hbar\xi(\xi+1)\nonumber\\
& &\ \ \times (T_{m+1}(\beta_{2}+\frac{i\hbar}{2})
\delta(\beta_{1}-\beta_{2}-i\frac{m+1}{2}\hbar)
-T_{m+1}(\beta_{2}-\frac{i\hbar}{2})
\delta(\beta_{1}-\beta_{2}+i\frac{m+1}{2}\hbar))\}
\end{eqnarray}
\noindent where 
\begin{eqnarray}
f_{1m}(\beta)=\frac{\Gamma(\frac{i\beta}{N\hbar}+1-\frac{1+m}{2N})
\Gamma(\frac{i\beta}{N\hbar}+1+\frac{1-m}{2N})
\Gamma(\frac{i\beta}{N\hbar}-\frac{\xi}{N}-\frac{1-m}{2N})
\Gamma(\frac{i\beta}{N\hbar}+\frac{\xi}{N}+\frac{1+m}{2N})}
{\Gamma(\frac{i\beta}{N\hbar}+1-\frac{\xi}{N}-\frac{1+m}{2N})
\Gamma(\frac{i\beta}{N\hbar}+1+\frac{\xi}{N}+\frac{1-m}{2N})
\Gamma(\frac{i\beta}{N\hbar}-\frac{1-m}{2N})
\Gamma(\frac{i\beta}{N\hbar}+\frac{1+m}{2N})}
\end{eqnarray}

\noindent {\bf Proof} Using the OPEs Eq.(11)-Eq.(13), we obtain that when 
Im$\beta_{2}<<$Im$\beta_{1}$
\begin{eqnarray*}
\Lambda_{l}(\beta_{1})
:\Lambda_{j_{1}}(\beta_{2}+i\frac{m-1}{2}\hbar)....
\Lambda_{j_{m}}(\beta_{2}-i\frac{m-1}{2}\hbar):
\end{eqnarray*}
\noindent is equal to 
\begin{eqnarray*}
f_{1m}(\beta_{2}-\beta_{1}):\Lambda_{l}(\beta_{1})
\Lambda_{j_{1}}(\beta_{2}+i\frac{m-1}{2}\hbar)....
\Lambda_{j_{m}}(\beta_{2}-i\frac{m-1}{2}\hbar):
\end{eqnarray*}
\noindent if $l=j_{k}$ for some $k\in \{1,....,m\}$ ; and 
\begin{eqnarray*}
& &f_{1m}(\beta_{2}-\beta_{1})\frac{
(i\frac{\beta_{2}-\beta_{1}}{N\hbar}
-\frac{\xi}{N}-\frac{1}{2N}+\frac{2k-m}{2N})
(i\frac{\beta_{2}-\beta_{1}}{N\hbar}
+\frac{\xi}{N}+\frac{1}{2N}+\frac{2k-m}{2N})}
{(i\frac{\beta_{2}-\beta_{1}}{N\hbar}
-\frac{1}{2N}+\frac{2k-m}{2N})
(i\frac{\beta_{2}-\beta_{1}}{N\hbar}
+\frac{1}{2N}+\frac{2k-m}{2N})}:\Lambda_{l}(\beta_{1})\\
& &\ \ \ \times 
\Lambda_{j_{1}}(\beta_{2}+i\frac{m-1}{2}\hbar)....
\Lambda_{j_{m}}(\beta_{2}-i\frac{m-1}{2}\hbar):
\end{eqnarray*}
\noindent if $j_{k}<l<j_{k+1}$. Here and below the case $l<j_{1}$ 
corresponds to $k=0$ and the case $l>j_{m}$ corresponds to $k=m$.
On the other hand,when Im$\beta_{2}>>$Im$\beta_{1}$,
\begin{eqnarray*}
:\Lambda_{j_{1}}(\beta_{2}+i\frac{m-1}{2}\hbar)....
\Lambda_{j_{m}}(\beta_{2}-i\frac{m-1}{2}\hbar):\Lambda_{l}(\beta_{1})
\end{eqnarray*}
\noindent is equal to 
\begin{eqnarray*}
f_{1m}(\beta_{1}-\beta_{2}):\Lambda_{j_{1}}(\beta_{2}+i\frac{m-1}{2}\hbar)
....\Lambda_{j_{m}}(\beta_{2}-i\frac{m-1}{2}\hbar)\Lambda_{l}(\beta_{1}):
\end{eqnarray*}
\noindent if $l=j_{k}$ for some $k\in\{1,...,m\}$ ; and 
\begin{eqnarray*}
& &f_{1m}(\beta_{1}-\beta_{2})\frac{
(i\frac{\beta_{1}-\beta_{2}}{N\hbar}
-\frac{\xi}{N}-\frac{1}{2N}-\frac{2k-m}{2N})
(i\frac{\beta_{1}-\beta_{2}}{N\hbar}
+\frac{\xi}{N}+\frac{1}{2N}-\frac{2k-m}{2N})}
{(i\frac{\beta_{1}-\beta_{2}}{N\hbar}
-\frac{1}{2N}-\frac{2k-m}{2N})
(i\frac{\beta_{1}-\beta_{2}}{N\hbar}
+\frac{1}{2N}-\frac{2k-m}{2N})}
:\Lambda_{j_{1}}(\beta_{2}+i\frac{m-1}{2}\hbar)\\
& &\ \ \ \times ....
\Lambda_{j_{m}}(\beta_{2}-i\frac{m-1}{2}\hbar)\Lambda_{l}(\beta_{1}):
\end{eqnarray*}
\noindent if $j_{k}<l<j_{k+1}$.Noting that 
\begin{eqnarray}
\lim_{\epsilon\longrightarrow0+}(\frac{1}{x+i\epsilon}-\frac{1}{x-i\epsilon})
=-2i\pi\delta (x)
\end{eqnarray}
\noindent we can obtain the commutation relations Eq.(17) for 
$T_{1}(\beta_{1})$ and $T_{m}(\beta_{2})$ after some 
straightforward caculation.Therefore, we 
complete the proof of Theorem 1 $\ \ {\large\bf \Box }$

In fact, the commutation relations for the generators of 
$\hbar$-deformed $W_{N}$ algebra have already been defined from   
the commutation relations of fundamental operators 
$\Lambda_{i}(\beta)$ in Eq.(6) and  the corresponding 
quantum Miura transformation in Eq.(7). So, one can also derive 
some similar commutation relations between  
$T_{i}(\beta)$ and $T_{j}(\beta)$ with $i,j>1$ 
using the same method as that in proof of the Theorem 1. 
These commutation relations are quadratic, and  
involve products of $T_{i-r}(\beta)$ and $T_{j+r}(\beta)$ with $r=1,...,
{\rm min}(i,j)-1$

In the case of N=2 ,this $\hbar$-deformed $W_{2}$ algebra becomes 
$\hbar$-deformed Virasoro algebra,which has been studied by us in the 
Ref.[12]. Here,we give an example for the case $N=3$.The generators of this 
case are  
\begin{eqnarray}
& &T_{1}(\beta)=\Lambda_{1}(\beta)+\Lambda_{2}(\beta)+\Lambda_{3}(\beta)\\
& &T_{2}(\beta)=
:\Lambda_{1}(\beta+\frac{i\hbar}{2})\Lambda_{2}(\beta-\frac{i\hbar}{2}):
+:\Lambda_{1}(\beta+\frac{i\hbar}{2})\Lambda_{3}(\beta-\frac{i\hbar}{2}) :
+:\Lambda_{2}(\beta+\frac{i\hbar}{2})\Lambda_{3}(\beta-\frac{i\hbar}{2}):
\end{eqnarray}
\noindent The commutation relations for these two generators are 
\begin{eqnarray}
& &f_{11}^{-1}(\beta_{2}-\beta_{1})T_{1}(\beta_{1})T_{1}(\beta_{2})
-f_{11}^{-1}(\beta_{1}-\beta_{2})T_{1}(\beta_{2})T_{1}(\beta_{1})
=-2i\pi \{i\hbar\xi(\xi+1)\nonumber\\
& &\ \ \times (T_{2}(\beta_{2}+\frac{i\hbar}{2})
\delta(\beta_{1}-\beta_{2}-i\hbar)
-T_{2}(\beta_{2}-\frac{i\hbar}{2})
\delta(\beta_{1}-\beta_{2}+i\hbar))\}\\
& &f_{12}^{-1}(\beta_{2}-\beta_{1})T_{1}(\beta_{1})T_{2}(\beta_{2})
-f_{12}^{-1}(\beta_{1}-\beta_{2})T_{2}(\beta_{2})T_{1}(\beta_{1})
=-2i\pi \{i\hbar\xi(\xi+1)\nonumber\\
& &\ \ \times (\delta(\beta_{1}-\beta_{2}-i\frac{3}{2}\hbar)
-\delta(\beta_{1}-\beta_{2}+i\frac{3}{2}\hbar))\}\\
& &f_{22}^{-1}(\beta_{2}-\beta_{1})T_{2}(\beta_{1})T_{2}(\beta_{2})
-f_{22}^{-1}(\beta_{1}-\beta_{2})T_{2}(\beta_{2})T_{2}(\beta_{1})
=-2i\pi \{i\hbar\xi(\xi+1)\nonumber\\
& &\ \ \times (T_{1}(\beta_{2}+\frac{i\hbar}{2})
\delta(\beta_{1}-\beta_{2}-i\hbar)
-T_{1}(\beta_{2}-\frac{i\hbar}{2})
\delta(\beta_{1}-\beta_{2}+i\hbar))\}
\end{eqnarray}
\noindent where the coefficient function $f_{ij}(\beta)$ are 
\begin{eqnarray*}
& &f_{11}(\beta)=
\frac{\Gamma(\frac{i\beta}{N\hbar}+1-\frac{1}{N})
\Gamma(\frac{i\beta}{N\hbar}+1)
\Gamma(\frac{i\beta}{N\hbar}-\frac{\xi}{N})
\Gamma(\frac{i\beta}{N\hbar}+\frac{\xi}{N}+\frac{1}{N})}
{\Gamma(\frac{i\beta}{N\hbar}+1-\frac{\xi}{N}-\frac{1}{N})
\Gamma(\frac{i\beta}{N\hbar}+1+\frac{\xi}{N})
\Gamma(\frac{i\beta}{N\hbar})
\Gamma(\frac{i\beta}{N\hbar}+\frac{1}{N})}\\
& &f_{12}(\beta)=\frac{\Gamma(\frac{i\beta}{N\hbar}+1-\frac{3}{2N})
\Gamma(\frac{i\beta}{N\hbar}+1-\frac{1}{2N})
\Gamma(\frac{i\beta}{N\hbar}-\frac{\xi}{N}+\frac{1}{2N})
\Gamma(\frac{i\beta}{N\hbar}+\frac{\xi}{N}+\frac{3}{2N})}
{\Gamma(\frac{i\beta}{N\hbar}+1-\frac{\xi}{N}-\frac{3}{2N})
\Gamma(\frac{i\beta}{N\hbar}+1+\frac{\xi}{N}-\frac{1}{2N})
\Gamma(\frac{i\beta}{N\hbar}+\frac{1}{2N})
\Gamma(\frac{i\beta}{N\hbar}+\frac{3}{2N})}\\
& &f_{11}(\beta)=f_{22}(\beta)
\end{eqnarray*}

\section{Screening currents}
In this section,we will consider the screening currents for the 
$\hbar$-deformed $W_{N}$ algebra.First,we introduce some zero mode operators. 
To each vector $\alpha\in{\bf P}$ (the $A^{(1)}_{N-1}$ type weight lattice 
defined in section 2.1),we associate operators $P_{\alpha}$, $Q_{\alpha}$ which  
satisfy
\begin{eqnarray}
[iP_{\alpha},Q_{\beta}]=<\alpha,\beta>\ \ \ ,\ \ \ (\alpha ,\beta \in {\bf P})
\end{eqnarray}
\noindent We shall deal with the bosonic Fock spaces $F_{l,k}$ 
$(l,k\in {\bf P})$ generated by $\lambda_{i}(-t)$ $(t>0)$ over the vacuum 
states $|l,k>$ .The vacuum states $|l,k>$ are defined by 
\begin{eqnarray*}
& &\lambda_{i}(t)|l,k>=0\ \ \ \ \ \ \ {\rm if}\ \ \ t>0\\
& &P_{\beta}|l,k>=<\beta,\alpha_{+}l+\alpha_{-}k>|l,k>\\
& &|l,k>=e^{i\alpha_{+}Q_{l}+i\alpha_{-}Q_{k}}|0,0>
\end{eqnarray*}
\noindent where $\alpha_{\pm}$ are some parameters related to $\xi$
\begin{eqnarray}
\alpha_{+}=-\sqrt{\frac{1+\xi}{\xi}}\ \ \ ,\ \ \  
\alpha_{-}=\sqrt{\frac{\xi}{1+\xi}}
\end{eqnarray}
\noindent and we also introduce $\alpha_{0}$ 
\begin{eqnarray}
\alpha_{0}=\sqrt{\xi(1+\xi)}
\end{eqnarray}
To each simple root $\alpha_{j}$ $(j=1,...,N-1)$, let us introduce two 
series bosons $s_{j}^{\pm}(t)$ which are defined by 
\begin{eqnarray}
& &s_{j}^{+}(t)=\frac{e^{j\frac{\hbar}{2}t}}{2sh\frac{\xi\hbar t}{2}}
(\lambda_{j}(t)-\lambda_{j+1}(t))\\
& &s_{j}^{-}(t)=\frac{e^{j\frac{\hbar}{2}t}}{2sh\frac{(1+\xi)\hbar t}{2}}
(\lambda_{j}(t)-\lambda_{j+1}(t))
\end{eqnarray}
\noindent By these simple root bosons, we can define the screening currents 
as follows
\begin{eqnarray}
& &S^{+}_{j}(\beta)=:exp\{-\int^{\infty}_{-\infty}s^{+}_{j}(t)e^{i\beta t}
dt\}:e^{-i\alpha_{+}Q_{\alpha_{j}}}\\
& &S^{-}_{j}(\beta)=:exp\{\int^{\infty}_{-\infty}s^{-}_{j}(t)e^{i\beta t}
dt\}:e^{-i\alpha_{-}Q_{\alpha_{j}}}
\end{eqnarray}
Then we have 

{\bf Theorem 2} The screening currents $S^{+}_{j}(\beta)$ satisfy 
\begin{eqnarray}
& &[:(e^{i\hbar\partial_{\beta}}-\Lambda_{1}(\beta))
(e^{i\hbar\partial_{\beta}}-\Lambda_{2}(\beta-i\hbar))....
(e^{i\hbar\partial_{\beta}}-\Lambda_{N}(\beta-i(N-1)\hbar)):
,S^{+}_{j}(\sigma)] \nonumber\\
& &=i\hbar(1+\xi)\{:(e^{i\hbar\partial_{\beta}}-\Lambda_{1}(\beta))....
(e^{i\hbar\partial_{\beta}}-\Lambda_{j-1}(\beta-i(j-2)\hbar))\nonumber\\
& & \ \ \times D_{\sigma ,i\hbar\xi}
(2\pi i\delta(\sigma -\beta -i\frac{j+\xi}{2}\hbar) A^{+}_{j}(\sigma))
\nonumber\\
& &\ \ \times e^{i\hbar\partial_{\beta}}(e^{i\hbar\partial_{\beta}}-
\Lambda_{j+2}(\beta-i(j+1)\hbar)).....
(e^{i\hbar\partial_{\beta}}-\Lambda_{N}(\beta-i(N-1)\hbar)):
\end{eqnarray}
\noindent with
\begin{eqnarray*}
A^{+}_{j}(\sigma)=:\Lambda_{j}(\sigma -i\frac{j+\xi}{2}\hbar)
S^{+}_{j}(\sigma):
\end{eqnarray*}
\noindent and the operator $D_{\sigma ,i\hbar\xi}$ is a diffenece operator 
with variable $\sigma$
\begin{eqnarray*}
D_{\sigma ,\eta}f(\sigma)\equiv f(\sigma)-f(\sigma+\eta)
\end{eqnarray*}
\noindent {\bf Proof} From the formulas Eq.(28)  ,we obtain the following 
commutation relations 
\begin{eqnarray*}
& &[\lambda_{j}(t),s^{+}_{j}(t')]=-\frac{2e^{\frac{t(1-j)}{2}\hbar}
sh\frac{(1+\xi)\hbar t}{2}}{t}\delta(t+t')\\
& &[\lambda_{j+1}(t),s^{+}_{j}(t')]=\frac{2e^{-\frac{t(1+j)}{2}\hbar}
sh\frac{(1+\xi)\hbar t}{2}}{t}\delta(t+t')\\
& &[\lambda_{j}(t),s^{+}_{l}(t')]=0\ \ \ \ ,\ \ \ {\rm if}\ \ |j-l|>1
\end{eqnarray*}
\noindent From these commutation relations and the formula Eq.(15) ,we can 
derive the following OPEs
\begin{eqnarray}
& &\Lambda_{j}(\beta_{1})S^{+}_{j}(\beta_{2})=
f^{+}_{jj}(\beta_{2}-\beta_{1}):\Lambda_{j}(\beta_{1})S^{+}_{j}(\beta_{2}):\\
& &S^{+}_{j}(\beta_{1})\Lambda_{j}(\beta_{2})=f^{+}_{jj}(\beta_{2}-\beta_{1})
:S^{+}_{j}(\beta_{1})\Lambda_{j}(\beta_{2}):\\
& &\Lambda_{j+1}(\beta_{1})S^{+}_{j}(\beta_{2})=
f^{+}_{j+1\ \ j}(\beta_{2}-\beta_{1}):\Lambda_{j+1}(\beta_{1})
S^{+}_{j}(\beta_{2}):\\
& &S^{+}_{j}(\beta_{1})\Lambda_{j+1}(\beta_{2})=
f^{+}_{j+1\ \ j}(\beta_{2}-\beta_{1}):
S^{+}_{j}(\beta_{1})\Lambda_{j+1}(\beta_{2}):\\
& &S^{+}_{j}(\beta_{1})\Lambda_{l}(\beta_{2})=
:S^{+}_{j}(\beta_{1})\Lambda_{l}(\beta_{2}):
=:\Lambda_{l}(\beta_{2})S^{+}_{j}(\beta_{1}):\ \ \ ,\ \ {\rm if}\ \ |j-l|>1
\end{eqnarray}
\noindent and 
\begin{eqnarray*}
& &f^{+}_{jj}(\beta)=\frac{\frac{i\beta}{N\hbar}-\frac{\xi}{2N}-\frac{1}{N}
+\frac{j}{2N}}
{\frac{i\beta}{N\hbar}-\frac{\xi}{2N}+\frac{j}{2N}
+\frac{\xi}{N}}\\
& &f^{+}_{j+1\ \ j}(\beta)=\frac{\frac{i\beta}{N\hbar}-\frac{\xi}{2N}
+\frac{1+\xi}{N}+\frac{j}{2N}}{\frac{i\beta}{N\hbar}-\frac{\xi}{2N}
+\frac{j}{2N}}
\end{eqnarray*}
\noindent The formula Eq.(37) means that 
\begin{eqnarray}
{\rm LHS\ \ of}\ \ {\rm Eq.}(32) = 
& &:(e^{i\hbar\partial_{\beta}}-\Lambda_{1}(\beta))....
(e^{i\hbar\partial_{\beta}}-\Lambda_{j-1}(\beta-i(j-2)\hbar))\nonumber\\
& &\ \ \times 
[(e^{i\hbar\partial_{\beta}}-\Lambda_{j}(\beta-i(j-1)\hbar))
(e^{i\hbar\partial_{\beta}}-\Lambda_{j+1}(\beta-i(j)\hbar)),
S^{+}_{j}(\sigma)]\nonumber\\
& &\ \ \ \times (e^{i\hbar\partial_{\beta}}-\Lambda_{j+2}(\beta-i(j+1)\hbar))
.....(e^{i\hbar\partial_{\beta}}-\Lambda_{N}(\beta-i(N-1)\hbar)):
\end{eqnarray}
\noindent Therefore, it is sufficient to consider the commutation relation 
\begin{eqnarray*}
[:(e^{i\hbar\partial_{\beta}}-\Lambda_{j}(\beta-i(j-1)\hbar))
(e^{i\hbar\partial_{\beta}}-\Lambda_{j+1}(\beta-i(j)\hbar)):,
S^{+}_{j}(\sigma)]
\end{eqnarray*}
\noindent According to the OPEs Eq.(33)---Eq.(36) ,we can derive that
\begin{eqnarray*}
[:\Lambda_{j}(\beta-i(j-1)\hbar)\Lambda_{j+1}(\beta-i(j)\hbar):
,S^{+}_{j}(\sigma)]=0
\end{eqnarray*}
\noindent Now we only need to consider the commutation relations between 
the term $\Lambda_{j}(\beta-i(j-1)\hbar)+\Lambda_{j+1}(\beta-i(j-1)\hbar)$ 
and the screening current $S^{+}_{j}(\sigma)$. From the OPEs Eq.(33)---Eq.(36)
, the formula  
Eq.(19), noting that
\begin{eqnarray*}
:\Lambda_{j}(\beta-i\frac{j-\xi}{2}\hbar)S^{+}_{j}(\beta+i\xi\hbar):
=:\Lambda_{j+1}(\beta-i\frac{j-\xi}{2}\hbar)S^{+}_{j}(\beta):
\end{eqnarray*}
\noindent and using the same method as that in the proof of Theorem 1, we 
have the following commutation relation  
\begin{eqnarray*}
[\Lambda_{j}(\beta-i(j-1)\hbar)+\Lambda_{j+1}(\beta-i(j-1)\hbar),
S^{+}_{j}(\sigma)]=i\hbar(1+\xi)D_{\sigma ,i\hbar\xi}
(2\pi i\delta(\sigma -\beta -i\frac{j+\xi}{2}\hbar) :A^{+}_{j}(\sigma)):
\end{eqnarray*}
\noindent Therefore, the Eq.(32) has been obtained. ${\bf \large \Box }$

Using the same method, we can have

{\bf Theorem} 3  The second series screening currents $S^{-}_{j}(\sigma)$ 
satisfy  
\begin{eqnarray}
& &[:(e^{i\hbar\partial_{\beta}}-\Lambda_{1}(\beta))
(e^{i\hbar\partial_{\beta}}-\Lambda_{2}(\beta-i\hbar))....
(e^{i\hbar\partial_{\beta}}-\Lambda_{N}(\beta-i(N-1)\hbar)):
,S^{-}_{j}(\sigma)] \nonumber\\
& &=i\hbar\xi\{:(e^{i\hbar\partial_{\beta}}-\Lambda_{1}(\beta))....
(e^{i\hbar\partial_{\beta}}-\Lambda_{j-1}(\beta-i(j-2)\hbar))\nonumber\\
& & \ \ \times D_{\sigma ,-i\hbar (1+\xi)}
(2\pi i\delta(\sigma -\beta +i\frac{-j+\xi +1}{2}\hbar) A^{-}_{j}(\sigma))
\nonumber\\
& &\ \ \times e^{i\hbar\partial_{\beta}}(e^{i\hbar\partial_{\beta}}-
\Lambda_{j+2}(\beta-i(j+1)\hbar)).....
(e^{i\hbar\partial_{\beta}}-\Lambda_{N}(\beta-i(N-1)\hbar)):
\end{eqnarray}
\noindent with
\begin{eqnarray*}
A^{-}_{j}(\sigma)=:\Lambda_{j}(\sigma +i\frac{-j+\xi +1}{2}\hbar)
S^{-}_{j}(\sigma):
\end{eqnarray*}
\noindent Therefore, the screening currents $S^{\pm}_{j}(\beta)$ commute 
with any $\hbar$-deformed $W_{N}$ algebra generators up to total difference.

\noindent{\bf Remark:} Let us take the conformal limit ($\hbar\longrightarrow 0$ and 
with $\xi$ and $\beta$ fixed ) ,the screening currents $S^{\pm}_{j}(\beta)$   
will become the ordinary screening current [8].

Theorem 2 and Theorem 3 imply that one can construct the intertwining 
operators (namely, vertex operators) for the $\hbar$-deformed $W_{N}$ 
algebra using the screening currents $S^{\pm}_{j}(\beta)$.In the next  
section we shall construct the vertex operators for the $\hbar$-deformed 
$W_{N}$ algebra.

\section{The vertex operators and its exchange relations}
In this section, we construct the type I and type II vertex operators 
for this $\hbar$-deformed $W_{N}$ algebra through the two series 
screening currents $S^{\pm}_{j}(\beta)$ .Firstly, we set
\begin{eqnarray}
\hat{\pi}_{\mu}=\alpha_{0}P_{\overline{\epsilon}_{\mu}}\ \ \ \  ,\ \ \ 
\hat{\pi}_{\mu\nu}=\hat{\pi}_{\mu}-\hat{\pi}_{\nu}
\end{eqnarray}
\noindent and 
\begin{eqnarray}
\hat{\pi}_{\mu\nu}F_{l,k} =<\epsilon_{\mu}-\epsilon_{\nu},
-(1+\xi)l+\xi k>F_{l,k}
\end{eqnarray}
\noindent Note that 
\begin{eqnarray}
e^{-i\alpha_{\pm}Q_{\gamma}}\hat{\pi}_{\sigma}e^{i\alpha_{\pm}Q_{\gamma}}
=\hat{\pi}_{\sigma}+\alpha_{0}\alpha_{\pm}<\sigma ,\gamma>
\end{eqnarray}
\noindent and this formula is very useful for caculating commutation 
relations of vertex operators. 
To each fundamental weight of $\omega_{j}$ $(j=1,....,N-1)$,let us introduce 
 two series bosons $a_{j}(t)$ and $a'_{j}(t)$ which are defined by 
\begin{eqnarray}
& &
a_{j}=\sum^{j}_{k=1}\frac{e^{-\frac{\hbar(j-2k+1)t}{2}}}
{2sh\frac{\hbar\xi t}{2}}\lambda_{k}(t)\ \ ,\ \ 
a'_{j}=\sum^{j}_{k=1}\frac{e^{-\frac{\hbar(j-2k+1)t}{2}}}
{2sh\frac{\hbar(1+\xi)t}{2}}\lambda_{k}(t)
\end{eqnarray}
\noindent and also define 
\begin{eqnarray}
U_{\omega_{j}}(\beta)=:exp\{\int^{\infty}_{-\infty}a_{j}(t)e^{i\beta t}dt\}:
e^{i\alpha_{+}Q_{\omega_{j}}}\ \ ,\ \ 
U'_{\omega_{j}}(\beta)=:exp\{\int^{\infty}_{-\infty}-a'_{j}(t)
e^{i\beta t}dt\}:e^{i\alpha_{-}Q_{\omega_{j}}} 
\end{eqnarray}
\noindent Because the vertex operators associated with each fundamental  
weight $\omega_{j}$ $(=2,...N-1)$ can be constructed from the 
skew-symmetric fusion of the basic ones $U_{\omega_{1}}(\beta)$  
and $U'_{\omega_{1}}(\beta)$ [1],it is sufficient to only 
deal with the vertex operators corresponding to the fundamental 
weight $\omega_{1}$.In order to caculate the exchange relations 
of the vertex operators,we first derive the following commutation  
relations
\begin{eqnarray*}
& &[a_{j}(t),s^{+}_{j}(t')]=-\frac{sh\frac{\hbar(1+\xi)t}{2}}
{tsh\frac{\hbar\xi t}{2}}\delta_{j,l}\delta (t+t')\ \ ,\ \ 
[a_{j}(t),s^{-}_{j}(t')]=-\frac{1}{t}\delta_{j,l}\delta (t+t')\\
& &[a'_{j}(t),s^{-}_{j}(t')]=-\frac{sh\frac{\hbar\xi t}{2}}
{tsh\frac{\hbar(1+\xi )t}{2}}\delta_{j,l}\delta (t+t')\ \ ,\ \ 
[a'_{j}(t),s^{+}_{j}(t')]=-\frac{1}{t}\delta_{j,l}\delta (t+t')\\
& &[a_{1}(t),a_{1}(t')]=-\frac{sh\frac{(N-1)\hbar t}{2}
sh\frac{(1+\xi)\hbar t}{2}}
{tsh\frac{N\hbar t}{2}sh\frac{\xi\hbar t}{2}}\delta(t+t')\\
& &[a_{1}(t),a'_{1}(t')]=-\frac{sh\frac{(N-1)\hbar t}{2}}
{tsh\frac{N\hbar t}{2}}\delta(t+t')\\
& &[a'_{1}(t),a'_{1}(t')]=-\frac{sh\frac{(N-1)\hbar t}{2}
sh\frac{\xi\hbar t}{2}}
{tsh\frac{N\hbar t}{2}sh\frac{(1+\xi)\hbar t}{2}}\delta(t+t')
\end{eqnarray*}
\noindent From the above relations, and taking the regularization in 
section 2.3, we can derive the following exchange 
relation
\begin{eqnarray}
& &U_{\omega_{1}}(\beta_{1})U_{\omega_{1}}(\beta_{2})=
r_{1}(\beta_{1}-\beta_{2})
U_{\omega_{1}}(\beta_{2})U_{\omega_{1}}(\beta_{1})\nonumber\\
& &U'_{\omega_{1}}(\beta_{1})U'_{\omega_{1}}(\beta_{2})=
r'_{1}(\beta_{1}-\beta_{2})U'_{\omega_{1}}(\beta_{2})
U'_{\omega_{1}}(\beta_{1})\nonumber\\
& &U_{\omega_{1}}(\beta_{1})U'_{\omega_{1}}(\beta_{2})=
\tau_{1}(\beta_{1}-\beta_{2})
U'_{\omega_{1}}(\beta_{2})U_{\omega_{1}}(\beta_{1})\nonumber\\
& &S^{+}_{j}(\beta_{1})S^{+}_{j+1}(\beta_{2})=-f(\beta_{1}-\beta_{2},0)
S^{+}_{j+1}(\beta_{2})S^{+}_{j}(\beta_{1})\nonumber\\
& &S^{-}_{j}(\beta_{1})S^{-}_{j+1}(\beta_{2})=-f'(\beta_{1}-\beta_{2},0)
S^{-}_{j+1}(\beta_{2})S^{-}_{j}(\beta_{1})\nonumber\\
& &U_{\omega_{1}}(\beta_{1})S^{+}_{1}(\beta_{2})
=-f(\beta_{1}-\beta_{2},0)S^{+}_{1}(\beta_{2})U_{\omega_{1}}(\beta_{1})
\nonumber\\
& &U'_{\omega_{1}}(\beta_{1})S^{-}_{1}=-f'(\beta_{1}-\beta_{2},0)S^{-}_{1}
(\beta_{2})U'_{\omega_{1}}(\beta_{1})\nonumber\\
& &U_{\omega_{1}}(\beta_{1})S^{-}_{1}(\beta_{2})=
-U_{\omega_{1}}(\beta_{1})S^{-}_{1}(\beta_{2})\ \ ,\ \ 
U'_{\omega_{1}}(\beta_{1})S^{+}_{1}(\beta_{2})=
-U'_{\omega_{1}}(\beta_{1})S^{+}_{1}(\beta_{2}) 
\end{eqnarray}
\noindent where the fundamental function $f(\beta ,w)$ and $f'(\beta ,w)$ 
(which play a very important role in constructing the vertex operators) are 
defined by 
\begin{eqnarray}
f(\beta ,w)=\frac{sin\pi(\frac{i\beta}{\hbar\xi}-\frac{1}{2\xi}-\frac{w}{\xi})}
{sin\pi(\frac{i\beta}{\hbar\xi}+\frac{1}{2\xi})}\ \ ,\ \ 
f'(\beta ,w)=\frac{sin\pi(\frac{i\beta}{\hbar(1+\xi)}+\frac{1}{2(1+\xi)}
+\frac{w}{1+\xi})}
{sin\pi(\frac{i\beta}{\hbar(1+\xi)}-\frac{1}{2(1+\xi)})}
\end{eqnarray}
\noindent and 
\begin{eqnarray}
& &r(\beta)=exp\{-\int^{\infty}_{0}\frac{2sh\frac{(N-1)\hbar t}{2}
sh\frac{(1+\xi)\hbar t}{2}shi\beta t}
{tsh\frac{N\hbar t}{2}sh\frac{\xi\hbar t}{2}}dt\}\\
& &r'(\beta)=exp\{-\int^{\infty}_{0}\frac{2sh\frac{(N-1)\hbar t}{2}
sh\frac{\xi\hbar t}{2}shi\beta t}
{tsh\frac{N\hbar t}{2}sh\frac{(1+\xi)\hbar t}{2}}dt\}\\
& &\tau (\beta)=\frac{sin\pi (\frac{1}{2N}-\frac{i\beta}{N\hbar})}
{sin\pi (\frac{1}{2N}+\frac{i\beta}{N\hbar})}
\end{eqnarray}
\noindent 
Now let us 
define the type I vertex operators $Z'_{\mu}(\beta)$ and the type II 
vertex operators $Z_{\mu}(\beta)$ $(\mu=1,...N)$ 
\begin{eqnarray}
& &Z_{\mu}(\beta)=\int_{C_{1}}\prod^{\mu -1}_{j=1}d\eta_{j}\ \ U'_{\omega_{1}}
(\beta)S^{-}_{1}(\eta_{1})S^{-}_{2}(\eta_{2})...S^{-}_{\mu-1}(\eta_{\mu-1})
\prod^{\mu-1}_{j=1}f'(\eta_{j}-\eta_{j-1},\hat{\pi}_{j\mu})\\
& &Z'_{\mu}(\beta)=\int_{C_{2}}\prod^{\mu -1}_{j=1}d\eta_{j}\ \ U_{\omega_{1}}
(\beta)S^{+}_{1}(\eta_{1})S^{+}_{2}(\eta_{2})...S^{+}_{\mu-1}(\eta_{\mu-1})
\prod^{\mu-1}_{j=1}f(\eta_{j}-\eta_{j-1},\hat{\pi}_{j\mu})
\end{eqnarray}
\noindent It is easy to see that the vertex operator $Z_{\mu}(\beta)$ and   
$Z'_{\mu}(\beta)$ are some bosonic operator intertwing the Fock 
spaces $F_{l,k}$
\begin{eqnarray}
& &Z_{\mu}(\beta)\ \ :\ \ F_{l,k}\longrightarrow 
F_{l,k+\overline{\epsilon}_{\mu}}\ \ ,\ \ 
Z'_{\mu}(\beta)\ \ :\ \ F_{l,k}\longrightarrow 
F_{l+\overline{\epsilon}_{\mu},k}
\end{eqnarray}
\noindent Here we set $\eta_{0}=\beta$ , the integration contour $C_{1}$ is 
chosen as :the contour corresponding to the integration variable $\eta_{j}$ 
enclose the poles $\eta_{j-1}+i\frac{\hbar}{2}-i\hbar\xi n(0\leq n)$ , and 
the other integration contour $C_{2}$ is chosen as : the contour corresponding 
to the integration variable $\eta_{j}$ enclose the poles 
$\eta_{j-1}-i\frac{\hbar}{2}-i(1+\xi)\hbar n(0\leq n)$ .

The constructure form of type I (type II) vertex operators of ours seems to 
be similar as that of vertex operators for $A_{N-1}^{(1)}$ face model given 
by Y.Asai et al [1],but with different bosonic operators and 
``coefficient parts" function $f(\beta ,w)$ $(f'(\beta ,w)$ .Thus the same 
trick [1]  can be used to caculate the commutation relations for our vertex 
operators. Using the method which was presented by Y.Asai et al in the 
apppendix B of the Ref.[1], we can derive the commutation relations 
for vertex operators $Z_{\mu}(\beta)$ and $Z'_{\mu}(\beta)$
\begin{eqnarray}
& &Z'_{\mu}(\beta_{1})Z'_{\nu}(\beta_{2})=\sum_{\mu '\nu '}
^{\overline{\epsilon}_{\mu}+\overline{\epsilon}_{\nu}  
=\overline{\epsilon}_{\mu '}+\overline{\epsilon}_{\nu '}}
Z'_{\mu '}(\beta_{2})Z'_{\nu '}(\beta_{1})\hat{W}'
\left(\begin{array}{ll}\hat{\pi}+\overline{\epsilon}_{\mu}+
\overline{\epsilon}_{\nu}&\hat{\pi}+\overline{\epsilon}_{\nu '}\\
\hat{\pi}+\overline{\epsilon}_{\nu}&\hat{\pi}
\end{array}|\beta_{1}-\beta_{2}\right)\\
& &Z_{\mu}(\beta_{1})Z_{\nu}(\beta_{2})=\sum_{\mu '\nu '}
^{\overline{\epsilon}_{\mu}+\overline{\epsilon}_{\nu}  
=\overline{\epsilon}_{\mu '}+\overline{\epsilon}_{\nu '}}
Z_{\mu '}(\beta_{2})Z_{\nu '}(\beta_{1})\hat{W}
\left(\begin{array}{ll}\hat{\pi}+\overline{\epsilon}_{\mu}+
\overline{\epsilon}_{\nu}&\hat{\pi}+\overline{\epsilon}_{\nu '}\\
\hat{\pi}+\overline{\epsilon}_{\nu}&\hat{\pi}
\end{array}|\beta_{1}-\beta_{2}\right)\\
& &Z_{\mu}(\beta_{1})Z'_{\nu}(\beta_{2})=
Z'_{\nu}(\beta_{2})Z'_{\mu}(\beta_{1})\tau (\beta_{1}-\beta_{2})
\end{eqnarray}
\noindent and the braid matrices 
$\hat{W}\left(\begin{array}{ll}
\hat{\pi}+\overline{\epsilon}_{\mu}+
\overline{\epsilon}_{\nu}&\hat{\pi}+\overline{\epsilon}_{\nu '}\\
\hat{\pi}+\overline{\epsilon}_{\nu}&\hat{\pi}
\end{array}|\beta\right)$ and $ \hat{W}'\left(\begin{array}{ll}
\hat{\pi}+\overline{\epsilon}_{\mu}+
\overline{\epsilon}_{\nu}&\hat{\pi}+\overline{\epsilon}_{\nu '}\\
\hat{\pi}+\overline{\epsilon}_{\nu}&\hat{\pi}
\end{array}|\beta\right)$ 
are some functions taking the value on operators $\hat{\pi}_{\mu\nu}$ like  
\begin{eqnarray}
& &\hat{W}'\left(\begin{array}{ll}
\hat{\pi}+2\overline{\epsilon}_{\mu}&\hat{\pi}+\overline{\epsilon}_{\mu}\\
\hat{\pi}+\overline{\epsilon}_{\mu}&\hat{\pi}\end{array}|\beta\right)
= r(\beta)\\
& &\hat{W}'\left(\begin{array}{ll}
\hat{\pi}+\overline{\epsilon}_{\mu}+\overline{\epsilon}_{\nu}&
\hat{\pi}+\overline{\epsilon}_{\nu}\\
\hat{\pi}+\overline{\epsilon}_{\nu}&\hat{\pi}\end{array}|\beta\right)
=-r(\beta)\frac
{sin\frac{\pi}{\xi}sin\pi(\frac{i\beta}{\hbar\xi}-
\frac{\hat{\pi}_{\mu\nu}}{\xi})}
{sin\pi(\frac{\hat{\pi}_{\mu\nu}}{\xi})sin\pi(\frac{i\beta}{\hbar\xi}+
\frac{1}{\xi})}\\
& &\hat{W}'\left(\begin{array}{ll}
\hat{\pi}+\overline{\epsilon}_{\mu}+\overline{\epsilon}_{\nu}&
\hat{\pi}+\overline{\epsilon}_{\mu}\\
\hat{\pi}+\overline{\epsilon}_{\nu}&\hat{\pi}\end{array}|\beta\right)
=r(\beta)\frac
{sin\pi(\frac{i\beta}{\hbar\xi})sin\pi(\frac{1}{\xi}+
\frac{\hat{\pi}_{\mu\nu}}{\xi})}
{sin\pi(\frac{\hat{\pi}_{\mu\nu}}{\xi})sin\pi(\frac{i\beta}{\hbar\xi}+
\frac{1}{\xi})}\\
& &\hat{W}\left(\begin{array}{ll}
\hat{\pi}+2\overline{\epsilon}_{\mu}&\hat{\pi}+\overline{\epsilon}_{\mu}\\
\hat{\pi}+\overline{\epsilon}_{\mu}&\hat{\pi}\end{array}|\beta\right)
= r'(\beta)\\
& &\hat{W}\left(\begin{array}{ll}
\hat{\pi}+\overline{\epsilon}_{\mu}+\overline{\epsilon}_{\nu}&
\hat{\pi}+\overline{\epsilon}_{\nu}\\
\hat{\pi}+\overline{\epsilon}_{\nu}&\hat{\pi}\end{array}|\beta\right)
=-r'(\beta)\frac
{sin\frac{\pi}{1+\xi}sin\pi(\frac{i\beta}{\hbar(1+\xi)}+
\frac{\hat{\pi}_{\mu\nu}}{1+\xi})}
{sin\pi(\frac{\hat{\pi}_{\mu\nu}}{1+\xi})sin\pi(\frac{i\beta}{\hbar(1+\xi)}-
\frac{1}{1+\xi})}\\
& &\hat{W}\left(\begin{array}{ll}
\hat{\pi}+\overline{\epsilon}_{\mu}+\overline{\epsilon}_{\nu}&
\hat{\pi}+\overline{\epsilon}_{\mu}\\
\hat{\pi}+\overline{\epsilon}_{\nu}&\hat{\pi}\end{array}|\beta\right)
=r'(\beta)\frac
{sin\pi(\frac{i\beta}{\hbar(1+\xi)})sin\pi(\frac{1}{1+\xi}+
\frac{\hat{\pi}_{\mu\nu}}{1+\xi})}
{sin\pi(\frac{\hat{\pi}_{\mu\nu}}{1+\xi})sin\pi(\frac{i\beta}{\hbar(1+\xi)}-
\frac{1}{1+\xi})}
\end{eqnarray}
\noindent Therefore, these braid matrices are not commute with the 
vertex operators and the exchange relations should be written in that order  
as Eq.(53) and Eq.(54).Noting that 
\begin{eqnarray*}
& &r(-\beta)|_{\xi\longrightarrow 1+\xi}=r'(\beta)\Delta_{N}(\beta)\\
& &\Delta_{N}(\beta)=\frac
{sin\pi(\frac{i\beta}{N\hbar}+\frac{1}{N})}
{sin\pi(\frac{i\beta}{N\hbar}-\frac{1}{N})}
\end{eqnarray*}
\noindent we find that the braid matrices 
$ \hat{W}\left(\begin{array}{ll}
\hat{\pi}+\overline{\epsilon}_{\mu}+
\overline{\epsilon}_{\nu}&\hat{\pi}+\overline{\epsilon}_{\nu '}\\
\hat{\pi}+\overline{\epsilon}_{\nu}&\hat{\pi}
\end{array}|\beta\right)$ and $ \hat{W}'\left(\begin{array}{ll}
\hat{\pi}+\overline{\epsilon}_{\mu}+
\overline{\epsilon}_{\nu}&\hat{\pi}+\overline{\epsilon}_{\nu '}\\
\hat{\pi}+\overline{\epsilon}_{\nu}&\hat{\pi}
\end{array}|\beta\right)$ are related to each other as follows
\begin{eqnarray*}
& & \hat{W}'\left(\begin{array}{ll}
\hat{\pi}+\overline{\epsilon}_{\mu}+
\overline{\epsilon}_{\nu}&\hat{\pi}+\overline{\epsilon}_{\nu '}\\
\hat{\pi}+\overline{\epsilon}_{\nu}&\hat{\pi}
\end{array}|-\beta\right)|_{\xi\longrightarrow 1+\xi}= 
\Delta_{N}(\beta)\hat{W}\left(\begin{array}{ll}
\hat{\pi}+\overline{\epsilon}_{\mu}+
\overline{\epsilon}_{\nu}&\hat{\pi}+\overline{\epsilon}_{\nu '}\\
\hat{\pi}+\overline{\epsilon}_{\nu}&\hat{\pi}
\end{array}|\beta\right) 
\end{eqnarray*}
\noindent When N=2,the factor $\Delta_{N}(\beta)=-1$,it occured  in studying  
the $\hbar$-deformed Virasoro algebra [12].
If both sides of Eq.(53) and Eq.(54) are acted on the special Fock 
space $F_{l,k}$ , noting that $l_{\mu\nu}$ and $k_{\mu\nu}$ 
\begin{eqnarray*}
l_{\mu\nu}=<\overline{\epsilon}_{\mu}-\overline{\epsilon}_{\nu} ,l>\ \ ,\ \ 
k_{\mu\nu}=<\overline{\epsilon}_{\mu}-\overline{\epsilon}_{\nu} ,k>
\end{eqnarray*}
\noindent are all some  
integer ,we have 
\begin{eqnarray}
& &Z'_{\mu}(\beta_{1})Z'_{\nu}(\beta_{2})|_{F_{l,k}}=\sum_{\mu '\nu '}
^{\overline{\epsilon}_{\mu}+\overline{\epsilon}_{\nu}  
=\overline{\epsilon}_{\mu '}+\overline{\epsilon}_{\nu '}}
Z'_{\mu '}(\beta_{2})Z'_{\nu '}(\beta_{1})|_{F_{l,k}}W'
\left(\begin{array}{ll}l+\overline{\epsilon}_{\mu}+
\overline{\epsilon}_{\nu}&l+\overline{\epsilon}_{\nu '}\\
l+\overline{\epsilon}_{\nu}&l
\end{array}|\beta_{1}-\beta_{2}\right)\\
& &Z_{\mu}(\beta_{1})Z_{\nu}(\beta_{2})|_{F_{l,k}}=\sum_{\mu '\nu '}
^{\overline{\epsilon}_{\mu}+\overline{\epsilon}_{\nu}  
=\overline{\epsilon}_{\mu '}+\overline{\epsilon}_{\nu '}}
Z_{\mu '}(\beta_{2})Z_{\nu '}(\beta_{1})|_{F_{l,k}}W
\left(\begin{array}{ll}k+\overline{\epsilon}_{\mu}+
\overline{\epsilon}_{\nu}&k+\overline{\epsilon}_{\nu '}\\
k+\overline{\epsilon}_{\nu}&k
\end{array}|\beta_{1}-\beta_{2}\right)
\end{eqnarray}
\noindent and 
\begin{eqnarray*}
& &W'\left(\begin{array}{ll}
l+2\overline{\epsilon}_{\mu}&l+\overline{\epsilon}_{\mu}\\
l+\overline{\epsilon}_{\mu}&l\end{array}|\beta\right)
= r(\beta)\\
& &W'\left(\begin{array}{ll}
l+\overline{\epsilon}_{\mu}+\overline{\epsilon}_{\nu}&
l+\overline{\epsilon}_{\nu}\\
l+\overline{\epsilon}_{\nu}&l\end{array}|\beta\right)
=r(\beta)\frac
{sin\frac{\pi}{\xi}sin\pi(\frac{i\beta}{\hbar\xi}+
\frac{l_{\mu\nu}}{\xi})}
{sin\pi(\frac{l_{\mu\nu}}{\xi})sin\pi(\frac{i\beta}{\hbar\xi}+
\frac{1}{\xi})}\\
& &W'\left(\begin{array}{ll}
l+\overline{\epsilon}_{\mu}+\overline{\epsilon}_{\nu}&
l+\overline{\epsilon}_{\mu}\\
l+\overline{\epsilon}_{\nu}&l\end{array}|\beta\right)
=r(\beta)\frac
{sin\pi(\frac{i\beta}{\hbar\xi})sin\pi(-\frac{1}{\xi}+
\frac{l_{\mu\nu}}{\xi})}
{sin\pi(\frac{l_{\mu\nu}}{\xi})sin\pi(\frac{i\beta}{\hbar\xi}+
\frac{1}{\xi})}\\
& &W\left(\begin{array}{ll}
k+2\overline{\epsilon}_{\mu}&k+\overline{\epsilon}_{\mu}\\
k+\overline{\epsilon}_{\mu}&k\end{array}|\beta\right)
= r'(\beta)\\
& &W\left(\begin{array}{ll}
k+\overline{\epsilon}_{\mu}+\overline{\epsilon}_{\nu}&
k+\overline{\epsilon}_{\nu}\\
k+\overline{\epsilon}_{\nu}&k\end{array}|\beta\right)
=r'(\beta)\frac
{sin\frac{\pi}{1+\xi}sin\pi(\frac{i\beta}{\hbar(1+\xi)}-
\frac{k_{\mu\nu}}{1+\xi})}
{sin\pi(\frac{k_{\mu\nu}}{1+\xi})sin\pi(\frac{i\beta}{\hbar(1+\xi)}-
\frac{1}{1+\xi})}\\
& &W
\left(\begin{array}{ll}
k+\overline{\epsilon}_{\mu}+\overline{\epsilon}_{\nu}&
k+\overline{\epsilon}_{\mu}\\
k+\overline{\epsilon}_{\nu}&k\end{array}|\beta\right)
=r'(\beta)\frac
{sin\pi(\frac{i\beta}{\hbar(1+\xi)})sin\pi(-\frac{1}{1+\xi}+
\frac{k_{\mu\nu}}{1+\xi})}
{sin\pi(\frac{k_{\mu\nu}}{1+\xi})sin\pi(\frac{i\beta}{\hbar(1+\xi)}-
\frac{1}{1+\xi})}
\end{eqnarray*}
It can be check that the Boltzmann weights 
$W\left(\begin{array}{ll}
c&d\\b&a\end{array}|\beta\right)$ and  
$W\left(\begin{array}{ll}
c&d\\b&a\end{array}|\beta\right)$ satisfy the star-triangle equations 
(or Yang-Baxter equation )
\begin{eqnarray}
\sum_{g}
& &W\left(\begin{array}{ll}d&e\\c&g\end{array}|\beta_{1}\right)
W\left(\begin{array}{ll}c&g\\b&a\end{array}|\beta_{2}\right)
W\left(\begin{array}{ll}e&f\\g&a\end{array}|\beta_{1}-\beta_{2}\right)
\nonumber\\
& &=\sum_{g}
W\left(\begin{array}{ll}g&f\\b&a\end{array}|\beta_{1}\right)
W\left(\begin{array}{ll}d&e\\g&f\end{array}|\beta_{2}\right)
W\left(\begin{array}{ll}d&g\\c&b\end{array}|\beta_{1}-\beta_{2}\right)
\end{eqnarray}
\noindent and unitary relation
\begin{eqnarray}
\sum_{g}W\left(\begin{array}{ll}c&g\\b&a\end{array}|-\beta\right)
W\left(\begin{array}{ll}c&d\\g&a\end{array}|\beta\right)=\delta_{bd}
\end{eqnarray}
\noindent Actually ,the Yang-Baxter equation Eq.(64) and the unitary relation 
Eq.(65) are direct results of the exchange relation of vertex operators in 
Eq.(62) and Eq.(63) (the associativity of algebra $Z_{\mu}(\beta)$ and 
$Z'_{\mu}(\beta)$ )

\noindent {\bf Remark}: In fact ,we have construct the bosonization of 
the vertex operators for the trigonometric SOS model of $A^{(1)}_{N-1}$ type.

\section*{Discussions}
We have constructed a $\hbar$-deformed $W_{N}$ algebra and its quantum  
Miura transformation.The $\hbar$-deformation of $W_{N}$ algebra can be 
obtained by two way: one can first derive the classical version of  
$\hbar$-deformed $W_{N}$ algebra  from studying the Yangian double with 
center $DY(\hat{sl}_{N})$ at the critical level as E.Frenkel and 
N.Reshetikin in studying the $U_{q}(\hat{sl}_{N})$ at critical level [10], 
then one construct the (quantum) $\hbar$-deformed $W_{N}$ algebra by 
quantizing the classical one ; Another way to construct the $\hbar$-deformed 
 $W_{N}$ algebra is by taking some scaling limit of q-deformed $W_{N}$ 
 algebra like Eq.(16).
Actually, the same phenomena also occur in studying Yangian double with 
center $DY(\hat{sl}_{N}$):the $DY(\hat{sl}_{N})$ can be considered as some  
scaling limit of $U_{q}(\hat{sl}_{N})$ algebra.

We only consider the $\hbar$-deformed $W_{N}$ algebra for generic $\xi$ .When 
$\xi$ is some rational number ($\xi=\frac{p}{q}\ \ , p$ and $q$ are two 
coprime integers), 
the realization of $\hbar$-deformed $W_{N}$ algebra in the Fock space 
$F_{l,k}$ would be highly reducible  and we have to throw out some states 
from the Fock space $F_{l,k}$ to obtain the irreducible component $H_{l,k}$.  
(Here we choose the same symbol as that in [1]). For N=2 ,the irreducible 
space $H_{l,k}$ (i.e $L_{l,k}$ in [12]) can be obtained by some BRST 
cohomology [12]. Unfortunately,the constructure of the BRST complex and the  
caculation of cohomology for $3\leq N$ is still an open problem . 

We also construct the vertex operators of type I and type II. 
These vertex operators satisfy some Fadeev-Zamolodchikov algebra with 
the face type Boltzmann weight as its constructure constant.In order to 
obtain the correlation functions as the traces of products of these vertex 
operators, we should introduce a boost operators $H_{1}$
\begin{eqnarray}
H_{1}=\int^{\infty}_{0}-\frac
{tsh\frac{N\hbar t}{2}sh\frac{\xi\hbar t}{2}}
{sh\frac{(N-1)\hbar t}{2}sh\frac{(1+\xi)\hbar t}{2}}a_{1}(-t)a_{1}(t)dt
\end{eqnarray}
\noindent which enjoys the property
\begin{eqnarray*}
e^{xH_{1}}a_{1}(t)e^{-xH_{1}}=e^{-xt}a_{1}(t)
\end{eqnarray*}
\noindent Morever,using the skew-symmetric fusion of N's vertex operators 
,we can obtain some invertibility for our vertex operators 
of the form like (3.19) and (c.20) in Ref.[1]. Then the correlation 
function can be described by the following trace 
function
\begin{eqnarray}
G(\beta_{1},....,\beta_{Nn})_{\mu_{1},....,\mu_{Nn}}=
\frac{tr(e^{-xH_{1}}Z'_{\mu_{1}}(\beta_{1})....Z'_{\mu_{Nn}}(\beta_{Nn}))}
{tr(e^{-xH_{1}})}
\end{eqnarray}

\end{document}